\documentclass[11pt,aaspp4]{aastex}

\slugcomment{Accepted to Astronomical Journal}

\shorttitle{Chemical Signatures} 
\shortauthors{Venn \etal}

\newcommand\kms{km\,s$^{-1}$}
\newcommand\afe{[$\alpha$/Fe] \,}

\newcommand\etal{{\rm et al.\,}}
\newcommand\eg{{\it e.g. }}
\newcommand\ie{{\it i.e. }}

\begin{document}


\title{Stellar Chemical Signatures And \\ Hierarchical Galaxy Formation} 

\author{Kim A. Venn\altaffilmark{1,2}}
\affil{Macalester College, Saint Paul, MN, 55105; 
       venn@clare.physics.macalester.edu}
\altaffiltext{1}{This work was completed while visiting
       the Institute of Astronomy, University of Cambridge, UK.}
\altaffiltext{2}{University of Minnesota, School of Physics \&
       Astronomy, 116 Church Street S.E., Minneapolis, 55455 }
\author{Mike Irwin}
\affil{Institute of Astronomy, University of Cambridge, Madingley Road,
       Cambridge, UK, CB3 0HA: mike@ast.cam.ac.uk}
\author{Matthew D. Shetrone}
\affil{McDonald Observatory, University of Texas at Austin;
       shetrone@astro.as.utexas.edu}
\author{Christopher A. Tout}
\affil{Institute of Astronomy, University of Cambridge, Madingley Road,
       Cambridge, UK, CB3 0HA: cat@ast.cam.ac.uk}
\author{Vanessa Hill}
\affil{Observatoire de Paris, GEPI and URA 8111 du CNRS, 92195 Meudon,
       France: Vanessa.Hill@obspm.fr}
\author{Eline Tolstoy}
\affil{Kapteyn Institute, University of Groningen, PO Box 800, 9700AV 
       Groningen, the Netherlands:  etolstoy@astro.rug.nl}

\begin{abstract}

To compare the chemistries of stars in the Milky Way dwarf spheroidal 
satellite galaxies (dSph) with stars in the Galaxy, we have compiled a 
large sample of Galactic stellar abundances from the literature.  
When kinematic information is available, we have assigned the stars to 
standard Galactic components through Bayesian classification based on 
Gaussian velocity ellipsoids. 
As found in previous studies, the \afe ratios of most stars in the 
dSph galaxies are generally lower than similar metallicity Galactic 
stars in this extended sample.   Our kinematically selected stars 
confirm this for the Galactic halo, thin disk, and thick disk components.
There is marginal overlap in 
the low \afe ratios between dSph stars and  Galactic halo stars 
{\it on extreme retrograde orbits} (V $< -420$ \kms), but this is
not supported by other element ratios.   Other element ratios
compared in this paper include r- and s-process abundances,
where we find a significant offset in the [Y/Fe] ratios that result
in a large overabundance in [Ba/Y] in {\it most} dSph stars compared
to Galactic stars.  
Thus, the chemical signatures of most of the dSph stars are distinct
from the stars in each of the kinematic components of the Galaxy.
This result rules out continuous merging of low mass galaxies similar 
to these dSph satellites during the formation of the Galaxy.
We do not rule out very early merging of low mass dwarf galaxies though, 
since $\le$1/2 of the most metal-poor stars ([Fe/H]$\le-1.8$) 
have chemistries that are in fair agreement with Galactic halo stars.    
We also do not rule out merging with higher mass galaxies, although 
we notice that the LMC and the remnants of the Sgr dwarf galaxy are 
also chemically distinct from the majority of the Galactic halo stars.
Formation of the Galaxy's thick disk by heating of an old thin disk
during a merger is also not ruled out, however the Galaxy's thick disk 
itself cannot be comprised of the remnants from a low mass (dSph) 
dwarf galaxy, nor a high mass dwarf galaxy like the LMC or Sgr, due 
to differences in chemistry.

The new and independent environments offered by the dSph galaxies also
allow us to examine fundamental assumptions related to the nucleosynthesis
of the elements.    The metal-poor stars ([Fe/H] $\le-1.8$)
in the dSph galaxies appear to 
have lower [Ca/Fe] and [Ti/Fe] than [Mg/Fe] ratios, unlike similar 
metallicity stars in the Galaxy.    Predictions from the $\alpha$-process 
($\alpha$-rich freeze out) would be consistent with this result if there
have been a lack of hypernovae in dSph galaxies.    The $\alpha$-process
could also be responsible for the very low Y abundances in the metal-poor 
stars in dSphs; since [La/Eu] (and possibly [Ba/Eu]) are consistent with 
pure r-process results, then the low [Y/Eu] suggests a separate r-process 
site for this light (first peak) r-process element.    We also discuss 
SNe II rates and yields as other alternatives though.  
In stars with
higher metallicities ([Fe/H]$\ge-1.8$), contributions from the s-process 
are expected; [(Y, La, \& Ba)/Eu] all rise as expected, and yet [Ba/Y]
is still much higher in the dSph stars than similar metallicity Galactic
stars.   This result is consistent with s-process contributions from
lower metallicity AGB stars in dSph galaxies, and is in good
agreement with the slower chemical evolution expected in the low mass 
dSph galaxies, relative to the Galaxy, such that the build up of metals 
occurs over much longer timescales.
Future investigations of nucleosynthetic constraints (as well as 
galaxy formation and evolution) will require an examination of many
stars within individual dwarf galaxies.

Finally, the Na-Ni trend reported by Nissen \& Schuster (1997) is 
confirmed in Galactic halo stars, but discuss this in terms of the 
general nucleosynthesis of neutron rich elements.  We do {\it not} 
confirm that the Na-Ni trend is related to the accretion of dSph 
galaxies in the Galactic halo. 

\end{abstract}

\keywords{
stars: abundances, kinematics,
Galaxy: abundances, stellar content, disk, halo,
galaxies: abundances, formation, dwarf,
Local Group.
}

\section{Introduction \label{intro}}

Modern cosmological models based on the Cold Dark Matter paradigm emphasize
the importance of hierarchical structure formation on all scales
(e.g., Navarro, Frenk, \& White 1997; White \& Rees 1978).
Galaxies like the Milky Way form as part of a local over-density in the
primordial matter distribution via the agglomeration of numerous smaller
building blocks which independently develop into dwarf galaxies.
By seeking the physical connections between local dwarf galaxies and the Milky 
Way we can therefore directly probe the formation and evolution of large 
galaxies, and in particular the extent to which merging with dwarf spheroidal
(dSph) galaxies could have affected the Galaxy.  

Proper motion surveys coupled with radial velocity and metallicity measures
have been used to look for and analyse the correlations between the 
orbital characterstics of stars and their overall metallicity measures 
(\eg Carney \etal 1996; Unavane, Wyse, \& Gilmore 1996).  
Together with the Hipparcos bright star catalogue 
which contains parallaxes, proper motions and radial velocities 
(Perryman \etal 1997), 
this has enabled classification of Galactic stars based on their kinematics 
alone and has identified groups of stars with, for example, disk or halo
kinematics, significant retrograde motion and those on highly elliptical 
orbits.  Some studies have specifically targeted retrograde stars to 
compare their abundances to prograde stars 
(\eg Stephens \& Boesgaard 2002; Ivan \etal 2003), 
while others have used statistical analyses of the kinematics 
of thousands of such stars to discover comoving groups; stars 
localized in dynamical phase space in the Galactic halo, 
which are possible signatures of accretion events 
(\eg Majewski \etal 1996; Helmi \etal 1999; Navarro \etal 2004).

Direct measurement of stellar abundances in Galactic satellite dwarf 
galaxies is a fairly new field of study which developed with the advent of
the 8-10m class telescopes and efficient high resolution spectrographs.  
Shetrone \etal (2001, 2003) determined the elemental abundances of 
36 red giants in seven dwarf galaxies and compared those to stars in the 
Galaxy, but found very little to connect these systems
(also see Tolstoy et al. 2003, Fulbright 2002, and 
Stephens \& Boesgaard 2002).
Stars in the dSph galaxies have lower \afe values in the mean than 
similar metal-poor Galactic halo stars, 
whereas the Galactic disk stars have much higher metallicities.
The chemical and kinematic analysis of 73 Galactic stars by 
Fulbright (2000, 2002) made partial use of kinematic information, 
and partial use of metallicity ([Fe/H] $< -1.0$), to separate the 
halo stars from thin and thick disk stars, which were then compared
to the (then available) averaged dSph stellar abundances.   
Alternatively, Stephens \& Boesgaard (2002) used kinematics alone 
to select 56 stars in various components of the Galactic halo (outer, 
intermediate, and high halo, and extreme retrograde stars), to compare
each of these to the (then available) stellar abundances in the 
Draco dSph galaxy.
Inspired by Stephens \& Boesgaard's work and Fulbright's work, 
we realized that it is now possible to compare the chemistry of 
{\it purely kinematically selected stars}\footnote{Stars are 
best sorted into their Galactic components based purely on
kinematics due to an overlap in the metallicity between the stellar 
components in the Galaxy (\eg Unavane \etal 1996, Chiba \& Beers 2000).}
in the Galactic halo, as well as the thin and thick disks, and compare 
each of these components to our larger dataset of stellar abundances, 
now in seven different dSph galaxies.   
Furthermore, our previously published dSph abundance 
comparisons did not contain many Galactic reference stars within the 
metallicity range of the dSph stars themselves, suggesting it would 
be worthwhile revisiting this issue.

In this paper, we use published kinematics and abundance measures of 
Galactic stars to update and improve the comparison between Galactic 
stellar populations and the dSph stellar abundances.  
We make use of the kinematic measures to statistically classify 
stars by their membership of the following Galactic components: 
halo, thick disk, thin disk, pronounced retrograde orbits and 
other high velocity stars.
We have adjusted the Mg and Ca abundances for four stars in the 
Scl dSph from Geisler \etal (2004) to account for differences in 
their adopted oscillator strengths (we raised their [Ca/H] by +0.05, 
and lowered their [Mg/H] by 0.18).  
We have also followed the suggestion by Shetrone (2004), and
homogenized the atomic data for spectral lines of Y, Ba, and Eu to 
improve the data quality and abundance comparisons from different authors
(discussed in Section~\ref{bay}).     We did not homogenize the
datasets for other elements though, particularly since different 
analyses use a variety of different spectral lines, often with little
or no overlap.   Superficial examinations do not reveal obvious offsets
or inconsistencies in the datasets we have combined for this paper for
most elements\footnote{Note that there is some discussion on the accuracy 
of the Ba abundances in our metallicity range 
(e.g., Johnson \& Bolte 2001; see Section~4). }, 
though more detailed inspection is certainly warranted.   
The effect of combining the datasets is likely to result in a larger 
spread in the abundances (and possibly abundance ratios) of order 0.1 
to 0.2 dex; this is the typical offset between recent analyses 
(discussed by Shetrone 2004), and is typical of the offsets in the
abundance comparisons between different model atmosphere techniques 
and atomic databases (examined by Shetrone \etal 2003). 
We do not attempt to homogenenize the errors analyses for all elements
from each of the Galactic analyses though, and instead adopt representative 
uncertainties
of $\Delta$[Fe/H] = $\pm$0.05, $\Delta$[{\sl X}/(Fe or H)] = $\pm$0.10,
and $\Delta$[{\sl X}/{\sl Y}] = $\pm$0.15 (where {\sl X, Y} represent
any element or combination of elements other than H and Fe).   
These minor caveats aside, the 
availability of large samples of stars with both detailed 
elemental abundance ratios and full kinematic information allows
us to improve upon the investigation of the galaxy-accretion connection 
through the combination of detailed elemental abundance ratios and a 
purely kinemetic classification scheme.

\section{Kinematics of Galactic Stars with Abundances }

We have examined the kinematic and abundance information in 
Fulbright (2000, 2002; 179 stars), Stephens \& Boesgaard (2002 = SB02;
41 stars not included in Fulbright), Hansen \etal (1998; 44 stars 
not in SB02 nor Fulbright),  Bensby \etal (2003; 21 thick disk and
45 thin disk stars), Nissen \& Schuster 
(1997 = NS97; 13 halo stars and 16 disk stars), 
Prochaska (2000; 10 thick disk stars),  
Reddy \etal (2003; 179 thin \& thick disk stars, not in the previous
analyses) and Edvardsson \etal (1993; 181 thin \& thick disk stars,
also not in previous analyses).     We have also included
elemental abundances from other surveys that do not have kinematic
information (see Table~1), including two unusual halo stars near
[Fe/H] $=-2$ by Ivans \etal (2003).    All data used in this analysis
is available electronically; a sample is shown in Table~2 . 

Where $U,V,W$ velocity information are available, we used a standard Bayesian
classification scheme to assign each data point to either: the thin disk, 
thick disk, or halo using the Galactic Gaussian velocity ellipsoid components 
given in Table~3.  Stars with extreme U,V,W measures (\ie a low probability
of being normal halo stars) were further subdivided into significantly 
retrograde motions ($V < -200$ \kms)\footnote {We use velocities corrected to
the Galactic reference frame throughout \ie. $(U,V,W)_\odot = (9,232,7)$}
and those with an extreme Toomre components ($T = \sqrt{U^2 + W^2} > 340$ \kms).
Since these samples of the solar neighbourhood are rather heterogeneous in 
nature and have been deliberately chosen to enhance the less populous local 
components, we used a uniform prior in assigning membership probabilities 
and have ignored subtleties of variation of the average rotation velocity 
with vertical distance from the Galactic Plane (\eg Chiba \& Beers 2000).
The resulting colour-coded distribution of the classified Galactic stars in 
the V-T plane is shown in Fig.~1 (left panel).   Also shown in Fig.~1 is
the variation of [Fe/H], \afe, and [Na/Fe] as a function of rotation 
velocity, which demonstrates the expected general trend of metallicity for 
the different Galactic components (similar to Fulbright's Fig.~5) and some
interesting general population trends in the element ratios.  
The metallicity plots also clearly show the large scatter and overlap 
between the different Galactic components (as discussed by 
Unavane \etal 1996, Gilmore \& Wyse 1998), most notably between
the halo and thick disk.

Kinematically, we are able to distinguish the thin disk, thick disk,
and halo stars, and confirm that there is a significant overlap in the 
chemical properties of these various components; see Fig.~1.   
We also highlight the significantly ($> 2\sigma$) retrograde 
and high Toomre velocity stars.  
Their metallicities overlap with the halo stars implying 
metallicity alone cannot be used to identify them.  
Intriguingly, the 
clump in metallicity of the majority of the extreme retrograde stars 
suggests an unexpected uniformity in their chemical properties, 
which is reflected in their \afe values and 
in mostly low [Na/Fe] values.  This suggests a common, and 
different origin for this population relative to the majority halo component.
These stars also differ from Ivans \etal (2003) three chemically peculiar (CP) 
outer halo stars; they have similar mean [Fe/H] values, but the CP stars
have more distinctive [(Mg, Ca, Ti, Ni)/Fe] ratio and [s\&r/Fe] upper limits.

For completeness we also performed a statistical decomposition of the 
elemental abundance patterns from Fulbright's (2000, 2002) large sample to
optimise the use of abundance information.  After examining the normalised 
parameter covariance matrix and undertaking a principal components analysis 
we concluded that:

(1) The $\alpha$-elements, including Mg, Si, Ca, and Ti, are tightly 
correlated ($\phi \approx 0.9$) with one another. The Eu abundances are also 
well correlated ($\phi \approx 0.75$) with these.  This is not unexpected 
since all are produced in type-II SNe events, although there are 
nucleosynthetic details that make an exact one-to-one correlation with Eu
unlikely (e.g., Sneden \etal 2000, Hill \etal 2002, Johnson \& Bolte 2002).   

(2) The next most dominant signal comes from [Na/Fe] which does not 
correlate well with \afe ($\phi \approx 0.0$) nor [Fe/H] ($\phi = 0.26$).  
Again this is not surprising.   Na is thought to form like the $\alpha$ 
elements (discussed further in Section~\ref{sodium}), yet the Na yield is 
controlled by the neutron excess.   The lack of correlation between [Na/Fe]
and \afe is unlike the results by Fulbright (2002). 

(3) The majority of the remaining variance (information) in the chemical 
ratios is fairly evenly distributed among the remaining elements with no
clear statistical signal standing out.

From this we conclude that the combination of probability-based kinematic 
classification with metallicity and abundance ratios offers a promising 
methodolgy for investigating 
stellar populations in the Galaxy, particularly as the sample sizes 
of HIP stars with abundances increases.   We investigate these 
correlations more closely in the following sections.

\section{Comparison of the \afe Ratios \label{alpha}}

The evolution of the chemical abundances in a galaxy is intimately 
linked to its star formation history (e.g., Tinsley 1979; Pagel 1998),
and one ratio of particular interest is \afe. 
Alpha-elements are produced primarily in high-mass stars of negligible 
lifetimes and ejected by SN II events, while 
iron is produced in both SN II and SN Ia events.  
A simple prediction is that stars that form shortly after the 
interstellar medium has been enriched by SNe II should have enriched 
\afe ratios, while those that form sometime after the SNe Ia contribute 
will have higher iron abundances and lower \afe ratios.    
As described by Matteucci (2003), the timescale for changes 
in the \afe ratio depends not only on the SFH, but also on the 
IMF, the SNe Ia timescale, and the timescales 
for mixing the SNe Ia and SNe II products back into the interstellar medium.
The \afe ratio observed in metal-poor stars in the Galactic halo 
is $\approx$+0.4, and includes O, Mg, Si, Ca, and Ti (as reviewed 
by McWilliam 1997).  This \afe ratio is consistent with an initial 
burst of star formation that had a standard IMF, after $\le$1 Gyr 
and before SNe Ia began to contribute significant amounts of 
Fe without $\alpha$-elements (Matteucci 2003).

In Fig.~2, the \afe ratios for three $\alpha$ elements (Mg, Ca, Ti) are
plotted individually and averaged together.   As expected, the metal-poor halo
stars in the Galaxy show high \afe ratios, while the thin disk stars 
show \afe ratios that approach solar with increasing metallicity.    
Notice also that the [$\alpha$/Fe] ratios of the thin disk blend smoothly
with those from the thick disk, which blend in smoothly with the halo stars.
Thus, the plateau in [$\alpha$/Fe] is the same, but 
the pattern in [$\alpha$/Fe] versus [Fe/H] differs. 
SB02 and Fulbright (2002) suggested that \afe is weakly and smoothly 
correlated with stellar apogalactic distance, or Galactic rest frame
velocity.  However, with our classifications based purely on
kinematics\footnote{ 
Fulbright's halo sample was selected partially by metallicity 
($-$2 $\le$ [Fe/H] $\le$ $-$1) which mixes his halo and disk populations.}, 
it is clear that these apparent correlations are primarily caused by 
{\it the low \afe ratios in the extreme retrograde stars alone}; see Fig.~1.   
Without extreme retrograde stars in the sample, there would be no 
significant \afe trend with distance, thus we do not support the
suggestion of an \afe trend in the Galactic halo.

We also find that the dSph stars are well separated from the majority 
of Galactic disk and halo stars in Fig.~2.   This was first shown by 
Tolstoy \etal (2003; Fig.~15).   On closer inspection, the extreme
retrograde component and possibly the high Toomre velocity component 
of the Galactic stars overlap with the dSph ratios better than the 
normal Galactic halo component. However these high velocity 
Galactic components do
not include stars with the lowest \afe ratios seen in the dwarf galaxies.   
Of course, from kinematics alone the extreme retrograde component is the 
most likely to come from satellite accretion events, unfortunately the
chemistry of these stars are not similar.    We will return to this
point in Section \ref{afedisc}.

Finally, in Fig.~2, 
we also notice that the scatter in \afe in the dSph 
stars is much larger for [Mg/Fe] than either [Ca/Fe] or [Ti/Fe].   
Thus, the overlap between the dSph and extreme retrograde (and possibly high 
Toomre velocity) stars is best for [Mg/Fe] and worst for [Ti/Fe].  
Recently, Shetrone (2004) has shown that the relative abundances of 
[Ca/Fe] and [Ti/Fe] in dSphs do tend to lie below those of [Mg/Fe],
which can also be seen in our Fig.~2.
This is not seen in most of the Galactic stars (except possibly in a
few very metal-poor stars, e.g., see Aoki \etal 2002), 
and could be related to differences in nucleosynthesis.    
For example, Mg forms through hydrostatic C- and O-burning in the 
cores of massive stars, whereas some isotopes of Ca 
(e.g., $^{44}$Ca, Woosley \& Weaver 1995) and Ti 
(e.g., $^{48}$Ti, Nakamura \etal 2001) 
are thought 
to form in the $\alpha$-process\footnote{The $\alpha$-process 
(or $\alpha$-rich freeze out) may occur when neutron-rich, 
$\alpha$-rich gas is out from nuclear statistical equilibrium, 
such as in the high entropy wind that blows from the surface of a proto 
neutron star following core collapse (Woosley \& Hoffman 1992;
Nakamura \etal 2001).}.   
If these processes occur together (in the same stars or in a fully
populated IMF) then the [Mg/Fe], [Ca/Fe], and [Ti/Fe] ratios will   
show the same trends as one another in metal-poor stars, such as
is seen in the Galactic halo stars.
However, these ratios may differ from one another in the dSph
galaxies due to differences in their star formation histories,
IMF, stellar yields, or mixing timescales.
  
Nakamura \etal (2001) suggest the $\alpha$-process occurs in hypernovae
(with energies $\ge$ 10$^{52}$ erg),
evidenced by high [Si/O] ratios in star burst galaxies.   The difference
in the [(Ca,Ti)/Fe] ratios compared to [Mg/Fe] in the dwarf galaxies may
suggest a lack of hypernovae in the dwarf galaxies.   This would be
a similar suggestion as that made by Tolstoy \etal (2003) of an
effectively truncated upper IMF in these low mass dwarf galaxies
with their slower chemical evolution and possibly lower total mass 
star formation events.
As another alternative, Regos \etal (2003) show that substantial amounts of
$^{44}$Ti, which decays to $^{44}$Ca, can be produced by SNe Ia through 
exploding He white dwarfs.
This source can help to explain the solar system $^{44}$Ca, which is
underestimated from the standard SN II yields (Timmes \etal 1995) and 
standard exploding Chandrasekhar-mass white dwarf yields 
(Livne \& Arnett 1995).   
If significant amounts of Ca and Ti are produced through standard
SN II, standard SN Ia, and exploding He white dwarfs, then differences
in the star formation histories in dSph galaxies {\it alone} 
could yield variations in [Ca/Mg] and [Ti/Mg] in dSphs relative to 
Galactic stars, without affecting the IMF.

\section{Comparison of s- \& r-process Ratios \label{bay}}

The \afe ratio is a fundamental chemical signature in stars because it 
depends on the relative contributions of SN II to SN Ia products available 
when the star formed.  Other element ratios are equally fundamental 
signatures, but possibly more difficult to interpret.
An example of another fundamental chemical signature are the s- and 
r-process ratios in a star, i.e., relative abundances of elements
formed through rapid (r-process) and slow (s-process) neutron captures.
Interpretation of these ratios is complicated by the uncertainty in 
the exact sources (and thus yields) of the r-process and s-process 
(c.f., Johnson \& Bolte 2002).
Generally, the r-process abundances are linked to SNe II events
and a high entropy neutrino wind at the surface of the newly formed
neutron star (Duncan \etal 1986; Woosley \etal 1994), 
although merging neutron stars (Rosswog \etal 2000) 
and asymmetric explosions and jet-like like outflows from
newly born neutron stars (LeBlanc \& Wilson 1970; Cameron 2003) 
are other possibilities. 
There are also two known sources for s-process elements,
including low-mass AGB stars responsible for the {\it main}
s-process (Busso \etal 1999), and helium burning in massive stars 
(the {\it weak} s-process) which only contributes material with 
$A < 90$ and hence elements lighter than or up to Zr
(e.g., Prantzos \etal 1990; Raiteri \etal 1992).

In the Galaxy, interpretations of the metal-poor stellar abundances
suggest that s-process contributions do not occur until 
[Fe/H] $\sim -2$ and are not significant until near [Fe/H] $\sim -1$ 
(e.g., Travaglio \etal 2004).  
Qian \& Wasserburg (2001) developed a phenomenological model 
that interprets all heavy element abundances in Galactic stars in 
terms of pure r-process contributions up to [Fe/H] $= -1.0$.  
However, it is not clear that the s- and r-process abundance 
ratios in the dwarf galaxies need to coincide with those in 
the Galaxy as a function of time or [Fe/H].  
The lower star formation efficiency in the dwarf galaxies 
(Tolstoy \etal 2003; Matteucci 2003) means that metals will build 
up more slowly with time than in the Galaxy, and thus we might expect 
contributions from more metal-poor stars (e.g., metal poor AGB stars)
at a given time or metallicity.
Coupled with the fact that stellar yields are often metallicity 
dependent , then the s- and r-process ratios in stars in dwarf galaxies 
may be different from comparable metallicity stars in the Galaxy.

To examine differences in the r- and s-process ratios in dSph galaxies,
we use the Y, Ba, La, and Eu abundances.   Y, Ba, and La are important 
because they sample different peaks in the neutron magic numbers.  
Y (Z=39) belongs to the first peak that builds through rapid captures 
around neutron magic number N=50.   Ba (Z=56) and La (Z=57) belong to 
the second peak that builds around N=82.   Unfortunately, we do not
sample any element from the third peak that builds around N=126.   Since 
these elements may form through either r- or s-processing, then we also
examine Eu, an element that is largely from the r-process in the
Sun (95\% Burris \etal 2000).
 
The ratios of [Y/Fe], [Ba/Fe], [La/Fe], and [Eu/Fe] in the dSphs and 
Galactic stars are shown in Fig.~3.   The atomic data and spectral lines 
of Y, Ba, and Eu were compared between the datasets.   Similar atomic data
were used for the BaII (1-5 lines) and EuII (1 line) abundances, while 
offsets were found for YII (1-6 lines).   Adjustments were made to 
the YII abundances to account for mean differences in adopted oscillator 
strengths, including adjusting the Nissen \& Schuster (1997) results 
down by $-0.09$ dex, adjusting the Stephens \& Boesgaard (2002) results
down by $-0.09$ dex, and adjusting the Edvardsson \etal (1993) results
down by $-0.19$ dex.   The BaII abundances in McWilliam (1995) were 
replaced with those from McWilliam (1998).   The EuII abundances for
the Edvardsson \etal (1993) dataset are from Koch \& Edvardsson (2002).
For LaII, we compare with Burris \etal (2000) and Johnson (2002) only.
The first impression is that the dSph
stars span a larger range in all three ratios at intermediate
metallicities than the Galactic stars.   At very low metallicities, the
large ranges seen in the Galactic star data are interpreted as inhomogeneous
mixing i.e., the proximity to a recent SN II event will vary the ratios 
of the r-process elements in newly formed stars (c.f., McWilliam 1997). 
Possibly inhomogeneous mixing plays a role at intermediate metallicities 
in the dSph stars, particularly as [Eu/Fe] shows a large range and is 
primarily an r-process element.   
This might be expected if their star formation was similar to 
the stochastic star formation observed in the Local Group dwarf 
irregular galaxies (e.g., Dohm-Palmer \etal 1997, 1998).
However, we also notice that [Y/Fe] is significantly lower/offset
in the dSph stars than in the Galaxy.   This includes roughly
half of the dSph stars, and suggests the r- and s-process enrichment 
of this element differs between the galaxies.

To examine the s-process and r-process contributions separately, 
we also examine 
the [Y/Eu], [Ba/Eu], and [La/Eu] ratios in Fig.~4.   Again, 
the Galactic stars show signs of inhomogeneous mixing in the 
most metal-poor stars (presumably due to variations in the 
{\it r-process} contributions to Y and Ba from localized SNe II events).
By [Fe/H] = $-2.0$, the [Ba/Eu] abundances are above the pure
r-process plateau (near [Ba/Eu] = $-$0.7, e.g., Burris \etal 2000),
however this is not clearly due to s-process contributions.  
The [La/Eu] abundances are close to the predicted pure r-process
values, and $^{57}$La should show similar behavior to $^{56}$Ba, 
having a similar nucleosynthetic history and s-process contributions 
in the solar system.
Johnson \& Bolte (2001) showed that the interpretation of the Ba
abundance in this metallicity range is complicated, and that the
rise in [Ba/Eu] may not be due to s-process contributions but 
difficulties in the analysis of the strong BaII 4554 \AA\ line.
Thus, the Galactic data would move closer to the pure r-process line,
but not the dSph data since we did not use the 4554 \AA\ line 
(Shetrone \etal 2003, 2001). 

The [Ba/Eu] ratios in dSph stars also suggest a smooth rise starting 
near [Fe/H] = $-2.0$.  This rise is expected 
to be due to contributions from the s-process from AGB stars. 
That [Y/Eu] remains low in most of the dSph stars suggests
contributions from {\it metal-poor} AGB stars, where first peak 
elements (like Y) are bypassed in favour of second peak (like Ba and La; 
and third peak, which is unobserved) elements because there are fewer 
nuclei to absorb the available neutrons.    
Thus, high [Ba/Y] ratios are consistent with the expectations from 
metal-poor AGB s-process yields (Travaglio \etal 2004).
We are not suggesting that all of the dSph stars show metal-poor AGB
contributions though.   The most metal-poor ([Fe/H]$\le-1.8$) 
dSph stars have [La/Eu] 
ratios in agreement with the pure r-process predictions\footnote{The 
only exception to this is UMi-K.   Despite its low metallicity 
([Fe/H]$ = -2.17$), this star has a very large [Ba/Eu] ratio suggestive 
of a signficant s-process contribution (unfortunately, we do not have 
[La/Eu] to confirm).   That its [Ba/Y] ratio is also very large is
in agreement with the expectations from metal-poor AGB contributions
outlined above. }
(and [Ba/Eu] is within 1~$\sigma$ of the pure r-process ratios).   
Therefore, the Y abundance in the most metal-poor dSph stars must also be 
purely r-processed.    Why is [Y/Eu] so low in these stars, 
and [Ba/Y] so high, if they are all from the same nucleosynthetic site?
This result suggests that the site of {\it r-processed} Y 
must differ from that of r-processed Ba, La, and Eu;
is there a {\it weak} r-process site (discussed further in Section 6.2)? 
Also, the source that produces Y in the metal-poor Galactic stars
must be absent in the dSphs or it must have a different time lag 
relative to the Ba, La, and Eu enrichments.

Regardless of the exact nucleosynthetic interpretation of the offsets
in the [Y/Eu] and [Ba/Y] ratios, it is clear from Fig.~4 that about
half of the dSph stars have lower [Y/Eu] ratios, and 2/3 have higher 
[Ba/Y] ratios than their Galactic counterparts, including the 
extreme retrograde halo stars.  
This offset suggests that no significant stellar component of
the Galaxy is formed from disrupted dSphs.   We discuss this 
further in Section~\ref{discussion}.

\section{The Na-Ni Relationship \label{sodium} }

Nissen \& Schuster (1997) noticed that a group of 8 halo stars have 
lower \afe ratios than disk stars of the same metallicity, and that these
stars were also underabundant in Na and Ni.   The [Na/Fe] ratios were
well correlated with [Ni/Fe], and they pointed out that the Na and 
$\alpha$-element depletions were largest for those stars probing the 
largest galactocentric distances\footnote{However, as mentioned in 
Section~\ref{alpha}, the NS97 sample probes stars with unusual kinematics, 
biased towards stars with nearly radial orbits and 8 $ < r_{apo} < $17.
Thus they do not sample stars that are truly in the outer Galactic halo.}.
This is in agreement with Fulbright (2002) who found low [Na/Fe] occured
only in stars with large galactocentric distances ($R_{max} > 20$ kpc). 
Stephens (1999) found no chemical differences ([Na/Fe] nor \afe) 
between 5 inner and 5 outer halo stars over a wider metallicity 
range ($-2.2 <$ [Fe/H] $< -1.0$) than NS97, however the larger
survey by Stephens \& Boesgaard (2002) did suggest a slight gradient 
in \afe (0.1 dex over 10 kpc, but with a dispersion of 0.1 to 0.2 dex),
but did not comment further on [Na/Fe] with distance.
NS97 further proposed that their stars might be remnants from an 
accreted dSph galaxy, and that perhaps the Na-Ni relationship they
detected is an indicator of the merging history of the Galaxy.

In Fig.~5 (top panel), we confirm the Na-Ni relationship found by NS97
for metal-rich stars ($-1.5 <$ [Fe/H] $< -0.5$; recall that we include 
the NS97 data, thus could expect to recover their Na-Ni relationship).  
The exact relationship appears slightly offset and steeper 
than reported by NS97 (the line in Fig.~5 is from their data alone).
The Na and Ni abundances in the dSph stars in this metallicity range 
also agree reasonably well with the NS97 trend 
(also discussed in Shetrone \etal 2003).
Broadening the metallicity range to
include all the stars from the literature examined here, then the correlation 
is less distinct, but still present for most of the Galactic stars
(see Fig.~5).  But, the dSph stars show no correlation.  

That the Na-Ni trend, reported by NS97, is {\it not} clearly seen 
in the dSph stars implies that the chemical evolution of these 
stars differs from the Galactic stars, as suggested by NS97, but 
not in a way that makes this a useful indicator of merging.
If the Na-Ni trend is a natural consequence of the nucleosynthesis 
of neutron-rich elements (as discussed below), then it is less likely
that a large fractions ($>$50\%) of the metal-rich ([Fe/H] $\sim -1.0$) 
halo may have formed through disrupted dSph galaxies like those studied 
here;  we suggested this in Shetrone \etal (2003) after observing 
that 50\% of our dSph stars with [Fe/H] $\sim -1.0$ show the Na-Ni 
relationship.
The low \afe ratios in the NS97 sample does indicate something odd 
in the formation of those stars though 
(discussed further in Section~\ref{afedisc}).
 
A slight and positive 
correlation between Na and Ni is a natural result of nucleosynthesis 
in massive stars (c.f., Clayton 1983, Section 5.6).   
As Timmes \etal (1995) suggests, Na is made by massive stars and 
delivered to the ISM in the SN II events.   The amount of Na 
produced is controlled by the neutron excess 
(primarily through the $^{22}$Ne content)\footnote{
The $^{23}$Na yield may depend on the initial heavy element 
abundance in a star (e.g., Arnett 1971), because the $^{22}$Ne 
{\it production} in a star is not strongly metallicity dependent
(Arnould \& Norgaard 1978).}
during hydrostatic C-burning.
While $^{23}$Na is naturally produced during hydrostatic C-burning 
in the core, the amount of $^{23}$Na produced relative to $^{24}$Mg
ranges from 1:2 to 1:5, depending on the temperature for C-burning 
(related to the mass of the star).  But what is noteworthy is that
$^{23}$Na is the only stable neutron-rich isotope produced in 
significant quantity during either the C- or O-burning stages. 
This is significant for the later formation of the stable $^{58}$Ni
isotope.   
During the SNe II event, the entire core photodissociates
the elements to protons and neutrons, which will be near equilibrium 
and recombine to form $^{56}$Ni, which beta decays to $^{56}$Fe, 
the dominant isotope of iron.   
Some $^{54}$Fe (primarily) and also $^{58}$Ni can be produced at this 
stage, depending on the abundance of the neutron-rich elements, 
e.g., $^{23}$Na, which is more common than other potential neutron-rich 
sources (such as $^{13}$C). 
The amount of $^{54}$Fe made is small compared to the total 
yield of iron (dominated by the $^{56}$Fe production), 
but this is the main source for $^{58}$Ni, the stable isotope of nickel.
In summary, the Ni production depends on
the neutron excess during the photodissociation of the core during the
SN II event, and the neutron excess will depend primarily on the
amount of $^{23}$Na produced earlier during hydrostatic C-burning.
Thus, a Na-Ni relationship is expected over a wide range in 
SNe II metallicities.

The contributions from SNe Ia can complicate the pattern, mainly because
of the variety of SNe Ia sources.
Nickel is produced without Na in the standard model of SNe Ia, where a
CO white dwarf tips over the Chandrasekhar limit and 
the C-burning region is completely photodissociated 
(Tsujimoto \etal 1995).   
This would flatten the Na-Ni correlation.
However, the route to SNe Ia is convoluted, and doubtless involves
rare binary star interactions (e.g., Reg\H{o}s \etal 2003).
It is quite uncertain what the Na yields from these might be, e.g.,  
if a CO white dwarf accretes a He white dwarf, perhaps the build 
up of He on the surface may undergo C-burning and release Na before 
the star photodissociates to form Ni in the SN event. 
Thus, it is not clear exactly what the SNe Ia contributions are to 
the Na-Ni correlation, although we can expect some sort of
slight, positive slope (where the specific slope depends on the 
SNe II yields and the variety of SNe Ia contributions), 
as seen in the Galactic data.

Finally, the fact that the dSph stars and very metal-poor halo 
stars (open symbols are those without kinematic information) 
lie well off of the trend
suggests two things; for the Na-rich stars, possibly there are
additional contributions to Na from merging white dwarfs, and
for the Ni-rich stars, possibly additional contributions from
a Ni rich SN II or SN Ia event.     That most of the Galactic stars
that lie well off the Na-Ni trend are the very metal-poor stars
supports this suggestion because there is evidence from the heavy
element ratios for inhomogeneous mixing at early epochs (McWilliam 1997).

\subsection{The Na-Mg Relationship:}
Before leaving our discussion on Na nucleosynthesis, we also find
the well known relationship between Na and Mg, as observed in
disk and halo stars (NS97, Hanson \etal 1998, Pilachowski \etal 1996).   
We confirm the Na-Mg relationship in the same metallicity
range examined by NS97 ([Fe/H] $> -1.3$) in dwarf stars, 
Fig.~6 (top panel), though with a slightly flatter slope.   
However, {\sl the giant stars} (bottom panel) 
{\sl in both the Galaxy and dSphs} show a scatter 
of [Na/Mg] abundances that is distinctly different from that 
of the dwarfs.   We believe this scatter is related to metallicity 
and not necessarily a problem in the Na determination in the giants.
A similar difference was noticed between field stars and cluster
giants by SB02, where the cluster abundances were discussed in terms
of internal, ``deep'' mixing which can alter the composition of
the stellar envelope through mixing with CNO-cycled gas.
Alternatively, here we notice that most of the (field) giants
have [Mg/H] $\le -1.5$ whereas most of the dwarfs are above this.  
The few giants above this metallicity are in good agreement with the
linear relation shown by the dwarfs, and the few dwarfs below this
metallicity are in good agreement with the large scatter seen in the
giants.  
Thus, we do not believe that this change in the Na-Mg pattern at 
[Mg/H] $\sim -1.5$ is due to a change in the nucleosynthesis of Na. 
Instead, it is more likely to be related to the quantities of Mg formed 
through hydrostatic C-burning (where $^{24}$Mg and $^{23}$Na
are both produced) relative to Mg formation through hydrostatic
O-burning and Ne-burning.   The Na-Mg correlation is tightest
at higher metallicities ($-1.5 <$ [Mg/H] $< -0.5$), thus the yields 
of Mg and Na are primarily from C-burning, 
whereas hydrostatic O- and Ne-burning were probably more
significant to the total Mg abundance at earlier epochs.

\section{Discussion \label{discussion}}

The purpose of this paper has been to compare the chemical properties
of stars in the current, expanded sample of low mass dwarf spheroidal 
galaxies to those of kinematically selected stars in the Galaxy.    
We have found that
no population of stars in the Galaxy is representative of stars in
the low mass dwarfs, in agreement with previous analyses
(Shetrone \etal 2001, 2003; Fulbright 2002; Stephens \& Boesgaard 2002; 
Ivans \etal 2003; Tolstoy \etal 2003).    Here, we discuss the differences 
in the fundamental chemical signatures, \afe and [(Ba/Y)/Eu], and point out
interesting differences that must occur in their nucleosynthetic 
histories.    Of course, we are comparing only a few stars from
seven dwarf galaxies, each with a unique star formation and evolutionary 
history.
The best way to approach further investigations for new, fundamental 
constraints of chemical nucleosynthesis is to examine many stars within
individual dwarf galaxies.

\subsection{dSph galaxies are not 
chemically similar to the Galaxy - I: \afe \label{afedisc} }

The \afe ratio has been proposed as a litmus test for chemical evidence 
of a satellite accretion in the Galactic halo (e.g., Unavane \etal 1996, 
Gilmore \& Wyse 1998).
Early comparisons of a sample of 13 Galactic halo stars by 
Nissen \& Schuster (1997) found a trend of decreasing \afe with
increasing apogalactic distance, $r_{apo}$.  They suggested that these 
outer halo stars might have been accreted from a dwarf galaxy with a slower 
chemical evolution history than the inner halo and disk stars.
These results were challenged by Stephens (1999) who compared 5 outer 
to 5 inner halo dwarf stars and found uniform abundances in all elements,
however a larger survey by Stephens \& Boesgaard (2002) suggest the 
potential for a slight trend in \afe versus galactocentric distance
(though the slope is $\le$ 1~$\sigma$ in the \afe dispersion).  
Fulbright (2002) also suggests that lower [Na/Fe] and \afe stars may
be found in the outer halo ($R_{max} > 20$ kpc). 
Interestingly the NS97 sample probes only a very limited range of the halo,
8 $ < r_{apo} < $17, and is biased toward stars which are on almost 
radial orbits.  This led Gilmore \& Wyse (1998) to question the conclusion 
of an accretion event origin for the low \afe NS97 stars, because their
very small $r_{peri}$ orbits would be 
difficult to populate from tidally disrupting a dwarf.  
Stephens (1999) concluded that the litmus test of low \afe 
ratios is probably insufficient to distinguish accreted debris 
from native stars. 

Our results from Fig.~2 suggest that low \afe is characteristic of
most stars in the dwarf galaxies, but should not be used as a litmus test.
Some dSph stars have higher \afe ratios like the Galactic stars,
and span a wide range in \afe from the halo plateau to subsolar values.  
In Fig.~7 (left panel), we show the \afe ratios for the stars in only
the Sculptor dSph (Shetrone \etal 2003 and Geisler \etal 2004). 
The stars with plateau \afe ratios have old ages ($\sim$15 Gyr from 
isochrone fitting; see Tolstoy \etal 2003), consistent with the
early chemical evolution of the halo (i.e., before significant SN~Ia 
contributions).   Stars with lower \afe ratios at higher metallicities
are consistent with theories of slow chemical evolution in dwarf galaxies 
(e.g., Matteucci 2003).   Sculptor's star formation history 
(e.g., Monkiewicz \etal 1999; Kaluzny \etal 1995), 
imply that the SN~Ia began to contribute 
near [Fe/H]=$-$1.8 (instead of $\sim -1$ like in the Galaxy).   
Notice that a merger of Sculptor at early times (before significant
SN~Ia contributions) would resemble normal (plateau \afe) halo stars.
 
On the other hand, the stars in Draco, also in Fig.~7 (right panel), 
show uniformly low \afe values at all metallicities, even though Draco 
is thought to have had a very similar star formation history to 
Sculptor (mostly old stars, Dolphin 2002).   
This may signal other effects in the chemical evolution
of Draco (such as blow-out or variations in its IMF), or there may simply
have too few stars in Draco to properly sample its abundance pattern
(the Draco \afe ratios are within 2 $\sigma$ of the higher halo values).
Since dSph's have a wide range of SFHs as determined from their CMDs
(e.g., Mateo 1998, Dolphin 2002), then it is likely there
is a large range in their \afe ratios, 
particularly at intermediate metallicities. 
This emphasizes the need for abundance ratios in many stars in 
individual dwarf galaxies to properly interpret their chemical
signatures and study halo formation through accretion events
(see Section~\ref{mergers}).

Does low \afe in a metal-poor star in the Galactic halo signal a 
formation site that differs from the rest of the halo? 
One Galactic halo star in our sample that clearly stands out in
Fig.~2 is BD+80\,245 (first noticed by Carney \etal 1997 from 
proper motion surveys); it is the Galactic halo star with very
low \afe (=$-0.1$) at [Fe/H] = $-$2.1 (abundances from Fulbright 
2000; also recently analysed in detail by Ivans \etal 2003).  
While its [$\alpha$/Fe] ratio does resemble that of a dSph star,
it is (nearly) unique amongst the stars presently studied
in the Galaxy (including when it is compared
to two other chemically peculiar stars in the outer Galactic halo
which have surprisingly high [Ti/Fe] ratios; Ivans \etal 2003).
Also, other elemental ratios
in this star do not resemble those in the dSphs (see Section 6.2).
The Nissen \& Schuster (1997) low \afe stars also stand out
in the Galactic data (in Fig.~2, they are the halo stars that
lie just below the thin disk near [Fe/H]=$-0.8$).   In fact,
the NS97 stars have [Fe/H] and \afe ratios that are similar
to stars in the LMC and Sgr dwarf galaxy (discussed in 
Section~\ref{mergers}).
Also, some globular clusters with low \afe abundances have 
been directly associated with merging galaxies, e.g., 
Pal\,12 (Dinescu \etal 2000; Brown, Wallerstein, \& Zucker 1997)
and Ter 7 (Sbordone \etal 2003; Ibata, Gilmore, Irwin 1995)
which are associated with the Sgr dSph galaxy.
Low \afe ratios may also be responsible for differences between 
photometric and CaT metallicity estimates for some clusters 
(e.g., Sarajedini \& Layden 1997), which might be attributed to 
globular clusters captured from accreted dwarf galaxies 
(or possibly from the Magellanic Clouds).
Of course, a very successful way to identify accretion debris 
is through the kinematic phase space constraints of the 
tidal stream debris (\eg Majewski \etal 1996; Helmi \etal 1999).
Determination of the chemical abundances in these stars could 
prove interesting!
However, the stars associated kinematically with Arcturus, and 
suggested by Navarro \etal (2004) to have an extragalactic 
origin, do {\it not} have low \afe ratios, making them unlike the 
stars in low mass dwarf galaxies.  In fact, the abundances in the Arcturus
group best resemble stars in the standard thick disk (discussed further
in Section~\ref{mergers}).

As a final note, Gratton \etal (2003) are currently reanalysing the 
stellar abundances
in many of the Galactic stars included here, as well as additional
stars, to put the stellar abundances onto a uniform metallicity scale.
Their initial results suggest that there is a larger 
dispersion in the \afe ratios in their ``counter-rotating'' stars,
stars with V $< 0$ \kms\ or V$_{LSR} < -200$ \kms\ 
($\ne$ extreme retrograde stars, as defined in this paper where
V$_{LSR} < -420$ \kms), 
however their samples are not kinematically separated.   
Their prograde stars are a mixture of 
halo, thin disk, and thick disk stars, while their counter-rotating 
stars are only halo stars 
(i.e., half of the Gaussian distribution in halo star velocities).  
Therefore, as expected, their counter-rotating stars have 
higher mean \afe and lower mean [Fe/H] than their prograde sample
(although the scatter in the \afe ratios for the counter-rotating
stars is much larger than for their prograde stars).

\subsection{dSph galaxies are not chemically 
 similar to the Galaxy - II: [Ba/Y] \label{ybadisc}} 

The offset in the [Ba/Y] ratios in the dSph stars shown in Fig.~4 
was previously seen by Shetrone \etal (2003).   This was interpreted 
as due to differences in the SFH in the dSphs, leading to AGB 
contributions in the dSphs from a more homogeneous sample of stars 
(mass and metallicity) compared to the Galaxy.  
In the Galaxy, a variety of AGB stars are likely to confuse any 
patterns in the individual AGB yields (McWilliam 1997). 
In Section~\ref{bay}, we developed this interpretation slightly.
For the dSph stars with [Fe/H] $\le -1.8$, the low Y suggests an 
{\it r-process} contribution from a different source than Ba and La, 
or a time delay in the {\it r-process} Y production. 
For the dSph stars with [Fe/H] $\ge -1.8$, the low Y suggests 
contributions primarily from low metallicity AGB stars.
While this interpretation is consistent with the theoretical predictions 
for metallicity-dependent AGB yields (Travaglio \etal 2004), and the 
differences in the dSph star formation histories, 
there are other possibilities to consider.
Other options that might affect the [Ba/Y] ratios in the chemical
evolution of a galaxy are a change in the SNe yields/rates, or
a change in the influence of the $\alpha$-process ($\alpha$-rich
freeze out). 

\noindent {\it SNe II Yields:} 
To investigate the role of SNe yields/rates, we examine 
the phenonenological model to reproduce the r-process 
abundances in the Galaxy by Qian \& Wasserburg (2001). 
Their model describes the synthesis of the r-process 
(and associated) elements in the Galaxy in terms of two 
types of SNe II, the H and L events.  The high frequency H 
events produce heavy r-process elements, like Eu, but no Fe 
(presumably retained in black hole remnants).  The low frequency L 
events produce light r-process elements (up to Ba) and Fe.   
Given the Eu and Fe abundance in a star with [Fe/H] $<-1$ 
in the Galaxy, then their model can accurately compute the 
rest of the r-process element ratios.
It is not clear that this model is appropriate for stars in the
dSph galaxies.   
The model predicts the low [Ba/Y] ratios found in the 
Galactic stars at these metallicities, but not the high [Ba/Y] 
ratios observed in the dwarf galaxy stars.   
Their model uses the observed discontinuity in the Galactic [Ba/Fe] 
ratio near [Fe/H]=$-2.5$ to determine the L and H SNe frequency and 
yields.  If this approach is correct, then there should be a similar 
behavior in the dSph stellar abundances, 
though not necessarily at the same metallicity due to differences 
in the mixing of metals in a galaxy, the L and H SNe frequencies,
and possibly in the star formation history.
We also note that they needed to 
correct the L event r-process contributions to Sr, Y, Zr, and Ba
from the Arlandini \etal (1999) estimates to reproduce the solar
r-process inventory. 
Surprisingly, if Arlandini \etal's  values are used, then
the predicted [Ba/Y] ratios are in good agreement with the dwarf
galaxy stellar ratios.    This ad hoc approach would imply
different contributions and/or sites for the formation of 
Sr, Y, Zr, and Ba between the dSph's and the Galaxy, or possibly
different retention yields if there is also mass loss driven by
the SNe II events.  

\noindent {\it SNe II Rates/IMF:} 
While changing the SNe II {\it yields} between sites is presently 
ad hoc, changing the frequency of events is less so, although it 
does suggest that the upper IMF is affected (and thus not universal).  
For example, if higher mass stars are primary sites for the 
production of Sr, Y, and Zr, then truncating the upper IMF would 
reduce their contributions in the dwarf galaxies.   This may even 
affect both the [Ba/Y] {\it and} \afe ratios since higher $\alpha$
yields are predicted from higher mass stars.  
This option was discussed in detail by Tolstoy \etal (2003). 
Smith \etal (2002) used this approach to explain their low [O/Fe] 
ratios in 12 LMC field giants.   Using a simplified chemical evolution model, 
they found one possible solution would be to lower the SN II rate slightly
more than the SN Ia rate per unit mass in the LMC.    
The young star forming complex R136 in the LMC is also an interesting 
example; Weidner \& Kroupa (2004) point out that a standard Salpeter 
IMF predicts that R136 should contain a star with M $\ge$ 750 M$_\odot$
(or at least 10-40 stars $\ge$150 M$_\odot$, the maximum mass observed). 
Does this suggest the upper IMF in R136 is truncated?   Weidner \& Kroupa
do conclude that there appears to be a maximum stellar mass limit
(e.g., the Eddington mass limit) or that the upper IMF is not invariant.
Nevertheless, if the upper mass limit is 150 M$_\odot$, then this is much 
higher than that needed to explain the low \afe ratios in the dSphs.  
Tolstoy \etal (2003) showed that the low \afe ratios in the dSph require 
a maximum stellar mass closer to 12 M$_\odot$. 
This low maximum stellar mass is problematic in terms of current
star formation theory.
If the IMF is a statistical property of a galaxy, then it should be fully 
populated over the integrated star formation history of the galaxy 
(C. Clarke and M.S. Oey, 2003, priv. communications). 
Detailed studies of star formation in galaxies show that
the concentrations of molecular gas where star formation occurs are highly
fractal, with structures from a few thousand to 1 M$_\sun$ (c.f., 
Larson review, 2003).   Even the Orion Nebula Complex (ONC), a site of
recent and active star formation, has a gas mass of only 10$^5$ M$_\odot$
and a total stellar mass of only $\sim$10$^3$ M$_\odot$, but it has formed 
a 50 M$_\odot$ star (e.g., Hillenbrand 1997). 
Thus, it is not clear whether the upper IMF differs between the dSph galaxies, 
and the star forming regions in the LMC and Galactic disk.   The molecular 
gas mass scales are significantly less than the total mass in all of
these systems, and the ONC suggests that even low gas mass systems can
form high mass stars. 

\noindent {\it The $\alpha$-Process:} 
The $\alpha$-process (or $\alpha$-rich freeze out) may occur when 
neutron-rich, $\alpha$-rich gas is out of nuclear statistical equilibrium, 
such as in the high entropy wind that blows from the surface of a proto 
neutron star following core collapse.  As mentioned in Section~\ref{alpha}, 
the  $\alpha$-process is thought to be important in the formation of 
$^{44}$Ca (Woosley \& Weaver 1995) and $^{48}$Ti (Nakamura \etal 2001).
It was also identified as a possible site for light r-process element 
formation by Woosley \& Hoffman (1992).
If the $\alpha$-process contributes to $\alpha$ {\it and} light r-process 
element production in the Galaxy, but {\it not} in the dwarf galaxies,
then we might expect lower \afe and lower [Y/Eu] ratios as is observed.
Since Nakamura \etal (2001) suggest the $\alpha$-process occurs in 
hypernovae (with energies $\ge$ 10$^{52}$ erg), 
then a lack of hypernovae in dwarf galaxies may be indicated. 
This would be similar to effectively truncating the upper IMF in dwarf 
galaxies.
If the $\alpha$-process is responsible for {\it both} the low 
[Mg/(Ca,Ti)] ratios (see Section 3) and the low [Y/Eu] observations
(see Section 4), then we might expect a relationship between
(Ca,Ti) and Y.   In Fig.~8, we plot the [Mg/Fe], [Ca/Fe], and [Ti/Fe]
against [Y/Fe] for metal-poor stars ([Fe/H] $\le -1.7$), which allows
us to focus on the SNe II yields alone.     The Galactic data 
does suggest a rise in [Ti/Fe] with [Y/Fe], which is marginally
supported by [Ca/Fe] and absent in [Mg/Fe] as expected from the 
$\alpha$-process predictions.
The extant dSph data does not show the same Galactic trends, 
in fact [Ti/Fe] may even decrease with increasing [Y/Fe].   
Thus, if hypernovae and the $\alpha$-process are responsible 
for a significant fraction of Ti in the Galaxy, then these plots 
suggest that this process is absent in the dSphs. 
Since our dSph stars are from a variety of galaxies, we suggest
this promising explanation of the $\alpha$-process in hypernovae 
(or lack of it) should be further investigated when many stars in 
each dSph galaxy become available. 

\subsection{What about hierarchical galaxy formation? \label{mergers}}

Regardless of the nucleosynthetic explanations for the low \afe and
high [Ba/Y] ratios in most stars in dSph galaxies, 
clearly the chemical signatures of these stars are not the same as
similar metallicity stars in the Galaxy.    Also, the chemical 
signatures of the stars in the dSphs are broadly similar to one 
another, suggesting that the differences in their chemical evolution
are small relative to the differences in the chemical evolution of
the Galaxy.  Together, these observations strongly imply that no 
significant component of the Galaxy formed primarily through the 
merger of galaxies similar to these low mass dSphs 
(i.e., after they began to form stars).     This is in agreement
with previous studies (Shetrone \etal 2001, 2003; Fulbright 2002;
Stephens \& Boesgaard 2002; Tolstoy \etal 2003; Ivans \etal 2003), 
but now includes detailed
examination of individual kinematic components including the thick disk, 
the thin disk, the halo, and halo stars with extreme retrograde orbits 
in the Galactic halo.     

We also have ages for the stars in the dSph galaxies.   
Tolstoy \etal (2003) used isochrones to find that most stars are
as old as the Galactic globular cluster fiducials.   Thus, the
Galaxy cannot have formed through late mergers of low mass dwarf 
galaxies similar to those seen today as predicted by some CDM models 
(e.g., Abadi \etal 2003).    However, we do not rule out significant
merging at early epochs.   Some ($\sim$1/2 to 1/3)
of the oldest stars in dSphs with [Fe/H] $\sim -2.0$ 
have the same chemical signatures as (standard)
Galactic halo stars, i.e., plateau [$\alpha$/Fe] and normal [Ba/Y]
and [Eu/Fe] ratios.    Also, any ancient merging of dwarf galaxies 
that occurs before the dwarf undergoes significant star formation 
would not be tracable by comparisons of stellar abundances.   This
scenario has been investigated by Bullock \& Johnston (2004)
where inner halo stars are formed from the gas deposits of accreted
low mass dwarf galaxies, thus inner halo stars do not resemble the 
stars in the surviving dSphs, but this does require that most of the
merging happened at the earliest epochs so that an insignificant
number of stars were formed {\it in the pre-merged satellites}.   Their
model also suggests significant substructure in the outer Galactic
halo such that outer Galactic halo stars will not resemble dSph
stars nor inner halo stars (as has been observed for three stars
by Ivans \etal 2003).
 
We do not rule out significant merging of higher mass dwarf galaxies
in our analysis.    Higher mass galaxies are unlikely to have had
the same slow chemical evolution of the low mass dSph's examined here.
Abundance ratios of stars in globular clusters in the LMC 
(Hill 2004; Hill \etal 2000) show Galactic halo-plateau values in [Mg/Fe] 
at metallicities up to [Fe/H] $\sim-1.7$.     Similarly,
Smecker-Hane \& McWilliam (2002) find higher \afe ratios in
three metal-poor stars ([Fe/H] $\sim-1.5$) in the Sgr dwarf 
remnant (although Bonifacio \etal 2004 question whether these 
stars are representative of the Sgr remnant or the globular cluster, 
M54).  These results overlap with stars in the 
Galactic halo and thick disk well\footnote{We note however that 
the LMC stars also show offsets such that low ratios are found for 
[O/Fe], [Ca/Fe], and [Ti/Fe], which are not well understood, and 
would be quite distinct from Galactic stars.}.   
  
The LMC and Sgr dwarf galaxy themselves cannot be used to explain 
the formation of the Galactic thick disk though, at least 
not through a late and violent merger event as envisioned by 
Abadi \etal (2003).
Differential abundances between thick and thin disk stars have 
shown that the thick disk stars have slightly {\it higher} \afe 
abundances than thin disk stars with [Fe/H] $\ge -0.6$
(Bensby \etal 2003; Feltzing \etal 2003).   While this cannot
be seen in the \afe ratios plotted for the whole dataset in 
Fig.~2, this can be seen beautifully when only the detailed 
and differential abundances from Bensby \etal are examined
(our Fig.~9, and their Figs. 12 and 13). 
While the simplest explanation for this chemical offset is
that the thick disk was enriched in SNe Ia products {\sl after} 
the formation of the thin disk (due to delayed Fe production in SN Ia),
it is not clear how this can happen.   
Bensby \etal summarize five observational constraints for the
discussion of the formation of the thick disk, and do favour a
merger scenario.   As they point out, heating from a merger 
event is expected to inflate the old thin disk to the velocity
dispersions of today's thick disk, however, the merging
galaxy would have had to be quite massive ($\sim0.1-0.2$ of
the Milky Way disk), and these galaxies are now rare in the 
Local Group.

We note however that the thick disk could not have formed from 
a disrupted galaxy that has similar chemistry to the 
LMC or the Sgr dwarf.   In Fig.~9, we show the [Mg/Fe] and 
[Ca/Fe] abundances from stars in the thick and thin disks, 
the dSphs, the LMC, and the Sgr dwarf.
The LMC and Sgr have {\it lower} \afe ratios than the thick disk
stars of comparable metallicity (Hill 2004; Hill \etal 2000, 1995; 
Luck \etal 1998; Bonifacio \etal 2004).   
Thus, if the thick disk formed through a merger, then the chemical 
properties of that merged galaxy were unlike any dwarf galaxy 
currently in the Local Group.
An alternative explanation might be that the thick disk formed through
a merger with a {\it gas-rich} dwarf galaxy that had undergone very 
little star formation.   The thick disk would be the remains of the 
heated old thin disk, and the present-day thin disk could have formed 
from diluted gas after settling into the rotational plane of the Galaxy. 
Dilution would lower [Fe/H], but have very little affect on 
the \afe ratios, and might fit the unusual abundance pattern.

Finally, it is well known that these seven surviving dSph galaxies 
may not represent the typical merging galaxy, particularly if their 
location in the Galaxy's dark matter halo has influenced their SFH, 
evolution, gas retention, 
structural properties, etc. (see the review by Mateo 1998).   
However, the stellar abundances in the LMC and Sgr also show 
lower \afe ratios than similar metallicity stars in the Galaxy 
(at least for [Fe/H] $\ge -1.0$).  
Stellar abundances in dIrr galaxies (gas rich, 
relatively isolated low mass galaxies in the outer parts 
of the Local Group) from young stars with [Fe/H]$\sim-1.0$
also show low \afe ratios (see Venn \etal 2001, 2003, 2004 and  
Kaufer \etal 2004 for results in NGC~6822, WLM, GR~8, and Sex~A).  
These abundances are in excellent agreement with results
from dSph galaxies, but not the Galaxy.  Stellar abundances
of old red giants in the isolated Local Group dIrr and dSph galaxies would
allow a direct determination of the chemistry in truly isolated galaxies,
but is not currently possible and will be an interesting program
for the next generation of ELT telescopes.
The assumption that the Galactic dSphs have always been in the dark matter
halo of our Galaxy is also not clear.  Proper motion surveys
suggest that most are bound to the Galaxy (e.g., Piatek \etal 2003;
Schweitzer \etal 1995), however, Fornax may have an unbound orbit
(Piatek \etal 2002).  In this case, a significant
fraction of the SFH of Fornax could have happened in isolation,
in an enviroment potentially similar to that of protogalactic fragments.

\section{Conclusions \label{conclusions}}

The purpose of this paper has been to compare the chemical properties
of stars in the current, expanded sample of low mass dwarf spheroidal 
galaxies to those of kinematically selected stars in the Galaxy.    
These comparisons can also be used to examine nucleosynthesis since
the chemical evolution of the dSph galaxies has been significantly
different from that of the Galaxy.    A summary of our conclusions
follows. 

(1) A Galactic sample of 781 stars from the literature, most (694)
    with UVW velocity information, can been assigned to the thin disk, 
    thick disk,
    standard halo, extreme retrograde motions (V $\le -420$ \kms) , 
    and extreme Toomre velocity components ($T = \sqrt{U^2 + W^2} > 340$ \kms)
    using Gaussian velocity ellisoid components.    The metallicity of
    these components has a significant overlap, therefore metallicity
    cannot be used generally to assign a star to a stellar population.
    The known extreme retrograde stars tend to have very similar 
    metallicities though, grouped near [Fe/H] $\sim -1.8$.

(2) The \afe ratios in the Galactic halo stars range from solar to
    the plateau value (\afe$\sim$0.4), with the extreme retrograde
    stars tending towards a middle value of \afe=0.2.   Similarly,
    the Galactic halo stars range from -0.4 $\le$ [Na/Fe] $\le$ 0.4,
    but with the extreme retrograde stars tending towards lower
    values ([Na/Fe]$\sim-0.5$).

(3) Comparison of the \afe ratios shows that no stellar population in 
    the Galaxy is representative of stars in the low mass dwarfs (in 
    agreement with previous analyses: Shetrone \etal 2001, 2003; 
    Fulbright 2002; Stephens \& Boesgaard 2002; Ivans \etal 2003; 
    Tolstoy \etal 2003).     The extreme retrograde stars have the
    closest \afe distribution to that of the dSph stars.
    It also appears that the individual $\alpha$-element ratios 
    in the dSphs show [Ti/Fe] $\le$ [Ca/Fe] $\le$ [Mg/Fe], at least 
    for the most metal-poor stars ([Fe/H] $\le -1.8$). 

(4) We find a new chemical signature in the dSph stars.  The [Ba/Y] and
    [Y/Eu] ratios suggest that $\sim$2/3 of the dSph stars are chemically
    different from stars in the Galaxy, {\it including the extreme 
    retrograde stars} which have normal Galactic halo ratios of these
    elements.

(5) Interpretation of the r- and s-process abundance ratios is 
    complicated.  In the most metal-poor ([Fe/H] $\le-1.8$) stars,
    we suggest that the offset between the dSph and Galactic halo
    stars could be due to an additional r-process site 
    (e.g., $\alpha$-process), which would be lacking in the dSphs.
    This is because the [La/Eu] ratios (and possibly the [Ba/Eu]
    ratios) are consistent with pure r-process abundances, and 
    therefore the [Y/Eu] should also be purely from the r-process.
    Thus, the offset in [Y/Eu] in the most metal-poor dSph stars
    is puzzling, unless there is an additional r-process site for
    this light (first peak) r-process element in the Galactic stars.

(6) In dSph and Galactic halo stars above [Fe/H]$=-1.8$, the rise
    in the s-process contributions to La, Ba, and Y can be seen.
    Thus, the high [Ba/Y] ratios seen in these stars also suggest
    that the s-process contributions to Ba and Y differ.   We suggest 
    that contributions from metal-poor AGB stars (Travaglio \etal 2003)
    in the dSph galaxies would be consistent with these observations and
    their different star formation histories (from that of the Galaxy). 
    Other possibilities are also discussed, such as an effectively 
    truncated IMF, and/or selective outflows of SNe II gas.

(7) We confirm the existance of the Ni-Na trend, as found by Nissen
    \& Schuster (1997) who also discussed this as a potential chemical 
    signature for stars from an accreted dwarf galaxy.    However, we
    suggest that this trend is a natural consequence of the nucleosynthesis 
    of neutron rich elements, and therefore {\it not} 
    related to the accretion of dSph galaxies in the Galactic halo. 

(8) We also discuss the [Na/Mg] relationship with [Mg/H] in terms of 
    yields from hydrostatic C-burning.    This relationship
    breaks down below [Mg/H]=$-1.5$, most likely due to Mg formation from 
    hydrostatic O- and Ne-burning at earlier epochs.    That most of
    the stars with [Mg/H] $\le -1.5$ are giants is not signficant in terms
    of nucleosynthesis; the giants are cooler and the spectral lines can be 
    sharper and stronger, thus sensitive to lower abundances.

Thus, the chemical signatures of the stars in the low mass dSph galaxies 
are unlike the available stars in all components of the Galaxy.   No 
component of the Galaxy (thick disk, thin disk, halo, extreme retrograde 
stars) could have formed primarily through the merger and disruption of 
low mass dwarf galaxies such as these at later epochs.    We do not rule
out merging with higher mass galaxies, though remnants of the Sgr dwarf 
galaxy found throughout the Galactic halo have retained a distinct chemical 
signature (e.g., Bonifacio \etal 2004; also from the globular clusters
associated with the Sgr remnant, Pal\,12 and Ter 7, Dinescu \etal 2000, 
Brown, Wallerstein, \& Zucker 1997, Sbordone \etal 2003, and 
Ibata, Gilmore, Irwin 1995).
We also do not rule out merging at very early epochs before low 
mass dwarf galaxies undergo signficant star formation.
Formation of the Galaxy's thick disk by heating of an old thin disk
during a merger is also not ruled out, however the Galaxy's thick disk 
itself cannot be comprised of the remnants from a low mass (dSph) 
dwarf galaxy, nor a high mass dwarf galaxy like the LMC or Sgr, due 
to differences in chemistry (Bensby \etal 2003; Prochaska 2000).
In the future, both the constraints for nucleosynthesis and the
testing of hierarchical galaxy formation through mergers with dSph-like
galaxies will require large samples of stars in individual dwarf galaxies.

\acknowledgements
Thanks to Mattheus Steinmetz for organizing the Thinkshop 2 in Potsdam,
Germany (June 2003), which stimulated this work.    
Special thanks to Bernard Pagel, Sean Ryan, Maria Lugaro, Yong Qian,
Chris Sneden, Bart Pritzl, 
and James Bullock for several helpful discussions and suggestions.
Also thanks to Verne Smith for making their Sculptor dSph stellar
abundances available. 
KAV would like to thank the NSF for support through a CAREER award, 
AST-9984073.
ET gratefully acknowledges support from a fellowship 
of the Royal Netherlands Academy of Arts and Sciences.  
CAT thanks Churchill College, University of Cambridge, for a fellowship.
MDS would like to thank the NSF for partial support from AST-0306884.
MJI thanks assorted breweries and wineries for help in maintaining sanity.

\clearpage
\epsscale{0.65}
\plotone{Venn.fig1.ps}
\figcaption{The upper left panel is a Toomre diagram showing stellar 
identifications using kinematic probabilities from velocity ellipsoids 
(see Table~3); thin disk (red), thick disk (green), halo (cyan).   
Two additional kinematic components stand out; 
an extreme retrograde component (black) and a 
high velocity Toomre component (blue). 
See Table~1 for data references.
The other panels show the variation of [Fe/H], [$\alpha$/Fe] and [Na/Fe] 
against rotation velocity, V, demonstrating the range of chemical variation 
for each component.   Notice the large scatter in the thick disk and halo
components, and especially the overlap in their metallicities.  
The extreme retrograde stars stand out both kinematically 
and as a function of chemical composition in all of these diagrams.
}

\clearpage
\epsscale{0.65}
\plotone{Venn.fig2.ps}
\figcaption{
[$\alpha$/Fe] versus metallicity for the individual $\alpha$ elements
Mg, Ca, Ti and the mean of the three using the same symbol coding as in 
Fig.~1.   We have also added stars without kinematic information as
hollow data points (all Galactic references in Table~1). 
The dSph stars (black squares; from Shetrone \etal 2003,
Shetrone \etal 2001, Geisler \etal 2004) clearly lay beneath 
most of the Galactic data for [Ca/Fe] and [Ti/Fe], with a wider spread
in [Mg/Fe].  The average dSph $\alpha$-index [(Mg+Ca+Ti)/3Fe] is significantly 
offset from the Galactic data. 
The mean of the extreme retrograde stars (black dots) generally 
lies between the mean of the dSph stars and the majority of the halo stars.
}

\clearpage
\epsscale{0.65}
\plotone{Venn.fig3.ps}
\figcaption{Ratios of [Y/Fe], [Ba/Fe], [La/Fe] and [Eu/Fe]. 
These sample elements from the first (Y) and second (Ba, La) 
r-process peak.    The [Y/Fe] offset in most of the dSph stars 
at all metallicities suggests differences in both the r- and 
s-process contributions (see text).   
Same symbols as in Figs.~1 and 2.
One dSph star, Scl-982 with [Fe/H]=$-0.97$ (Geisler \etal 2004), 
is not seen on these plots because each [{\sl X}/Fe] ratio lies 
above the plot axes; similarly, the dSph star UMi-K with 
[Fe/H]=$-2.17$ (Shetrone \etal 2001) 
has a [Ba/Fe] ratio that lies above the plot axis.
}

\clearpage
\epsscale{0.65}
\plotone{Venn.fig4.ps}
\figcaption{Ratios of [Y/Eu], [Ba/Eu], [La/Eu] and [Ba/Y].   
These ratios are used to examine s-process enrichments. 
The [La/Eu] ratios suggest there is no significant s-process
enrichment until [Fe/H] $\sim-1.8$ in both the Galactic and
dSph stars ([Ba/Eu] should also show this, but may be compromised
as discussed in the text).   That [Y/Eu] is clearly lower in the 
dSph stars at all metallicities suggests differences in both
the r- and s-process contributions (see text).   
The pure r-process estimates from solar system abundances are shown
from Arlandi \etal (1999; dashed line) and Burris \etal (2000; dotted
line).  Same symbols as in Figs.~1 and 2. 
}

\clearpage
\epsscale{0.5}
\plotone{Venn.fig5.ps}
\figcaption{
The relationship between [Na/Fe] and [Ni/Fe].
In the top panel, a narrow metallicity range
($-1.5 <$ [Fe/H] $< -0.5$) is shown, which clearly
shows a correlation as first noticed by NS97  
(solid line is directly from NS97). 
In the bottom panel, there are no restriction in 
metallicity.   The correlation is less distinct, but 
still present in the Galactic stars.  We suggest that
a slight and positive correlation between Na and Ni
is a natural result of nucleosynthesis in massive stars.
Same symbols as in Figs.~1 and 2.
}

\clearpage
\epsscale{0.5}
\plotone{Venn.fig6.ps}
\figcaption{
Na and Mg abundances in Galactic dwarfs (top panel) using the same 
symbols as in Figs. 1 and 2.  NS97 (supported by Hanson \etal 1998, 
Stephens 1999, SB02) found a tight correlation 
between the Na and Mg abundances for [Mg/H] $> -1.0$ 
(solid line is from their paper).
We confirm the tight correlation between Na and Mg and find that it 
extends to [Mg/H] $\sim -1.5$.   We suggest that above this metallicity
the yields of both $^{\rm 24}$Mg and $^{\rm 23}$Na are correlated
through their mutual production in hydrostatic C-burning.
However, the Na-Mg relationship is absent in the Galactic (and dSph) 
giants (bottom panel).  Most of these giant stars are more metal-poor 
(and cooler) than most of the dwarf stars.   Both dwarfs and giants 
with [Mg/H] $\le -1.5$ show a large scatter in the Na abundances, 
which we suggest is due to variations in their productions 
(e.g., more significant contributions to Mg from hydrostatic 
O- and Ne-burning) and possibly due to variations 
in the neutron seed elements at low metallicities (needed for Na 
production; see text). 
}

\clearpage
\plotfiddle{Venn.fig7.ps}{6.0cm}{0}{150}{150}{-550}{0}
\figcaption{
The \afe ratios for Galactic stars from Fig.~2 compared to
the stellar abundances in Sculptor (left panel; data from Shetrone \etal 2003
and Geisler \etal 2004) and Draco (right panel; data from SCS01). 
The most metal-poor stars in Sculptor have overlapping \afe with
the Galactic halo stars, however the higher metallicity stars have 
very low \afe, probably indicating SN Ia contamination starting at 
lower metallicities than in the Galaxy.  The dashed line is a linear
regression of the Sculptor data.
No stars in Draco have high \afe ratios though 
(the same dashed line from the Sculptor data is shown for comparison).
The low \afe ratios
in the most metal-poor stars in Draco suggest may indicate 
selective ($\alpha$-element) SN II blow out or a truncated upper IMF. 
Same symbols as in Fig.s~1 and 2.
}

\clearpage
\includegraphics*[scale=0.65]{Venn.fig8.ps}
\figcaption{Ratios of [Y/Fe] versus the alpha-elements, 
([Mg/Fe], [Ca/Fe], [Ti/Fe]) for metal-poor stars ([Fe/H]$\le-1.7$).   
The Galactic abundances appear to show a correlation between
Ti (and possibly Ca) with Y that may be related to the $\alpha$-process
($\alpha$-rich freeze out; see text).   A correlation is not
seen in the dSph data, which may suggest a lack of the 
$\alpha$-process source, such as hypernovae (Nakamura \etal 2001). 
Same symbols as in Figs.~1 and 2.
}

\clearpage
\epsscale{0.55}
\plotone{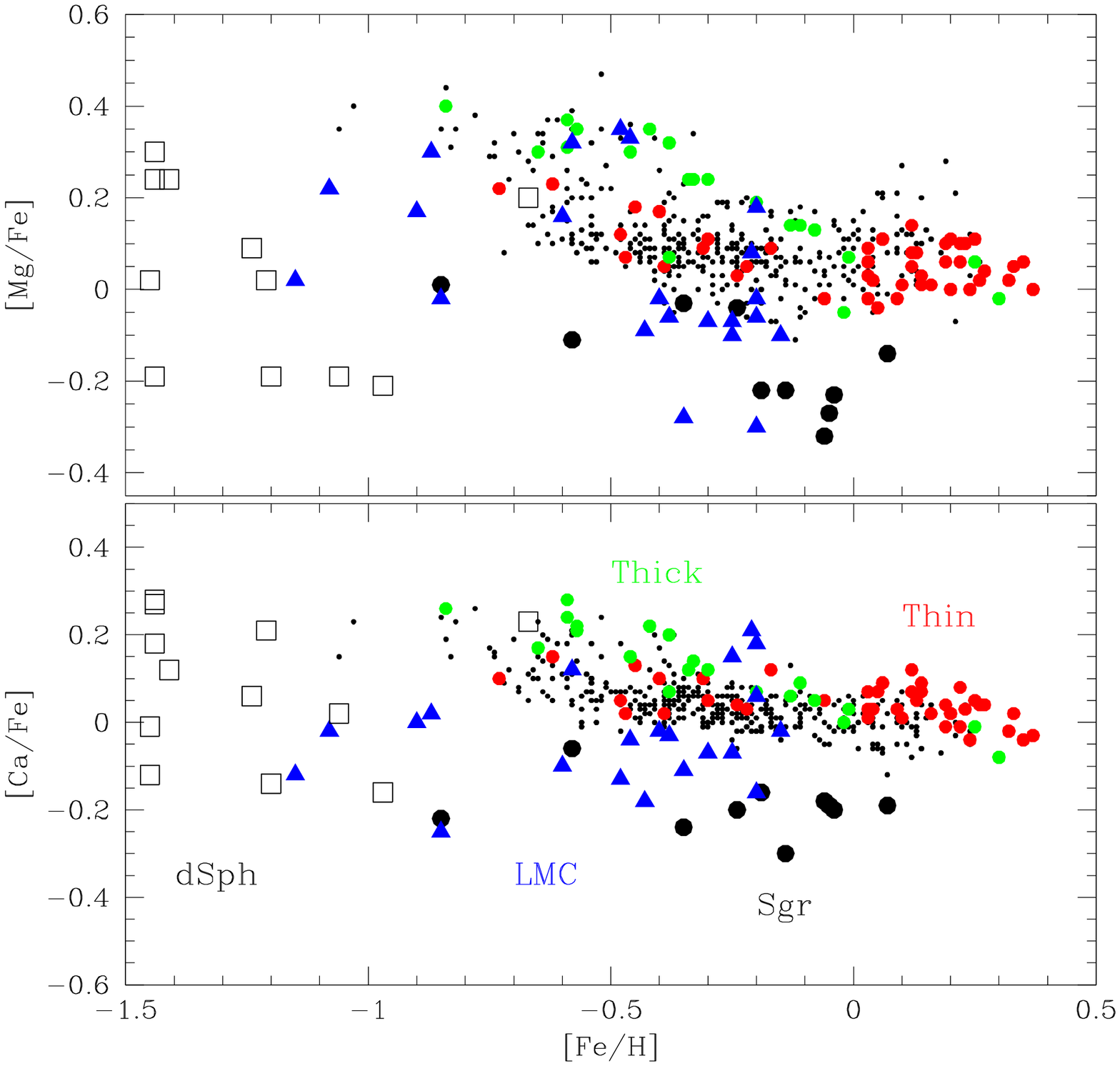}
\figcaption{[Mg/Fe] and [Ca/Fe] for stars in the Galaxy
versus those for stars in the LMC, dSphs, and Sgr dwarf.
Thick disk (green; Bensby \etal 2003), thin disk 
(red; Bensby \etal 2003), other disk stars (small black dots;
Edvardsson \etal 1993, Reddy \etal 2003) are compared with 
those in the LMC (blue triangles; Hill \etal 2003, 1995; Luck \etal 1998), 
the Sgr dwarf (large black dots; Bonifacio \etal 2004), 
and the dSphs (black squares; Shetrone \etal 2003, 
Geisler \etal 2004, Shetrone \etal 2001).
}


\clearpage
\begin{deluxetable}{lccl} 
\footnotesize
\tablecaption{Data Sources \label{datacatalogue}}
\tablewidth{0pt}
\tablehead{
\colhead{Reference} & \colhead{No. } & 
\colhead{[$\alpha$/Fe]}  & \colhead{Comment} 
} 
\startdata
                          &  &        & {\it With Kinematics} \\
Fulbright 2000, 2002      & 179 & [(MgI+CaI+TiI)/3FeI] &  thin, thick, halo \\
Stephens \& Boesgaard 2002 & 39  & [((MgI+CaI)+(TiI+TiII)/2)/3FeI] &  thin, thick, halo \\
Bensby \etal 2003         & 66 & [((MgI+CaI+(TiI+TiII)/2)/3FeI] & thin, thick \\
Nissen \& Schuster 1997   & 21 & [(MgI+CaI+TiI)/3FeI] &  thin, thick, halo \\
Hanson et al. 1998        & 19 & [(MgI+CaI)/2FeI] &  thin, thick, halo          \\
Prochaska et al. 2000     & 10 & [((MgI+CaI)+(TiI+TiII)/2)/3FeI] &  thick \\
Reddy et al. 2003        & 179 & [(MgI+CaI+TiI)/3FeI] & thin, thick \\
Edvardsson et al. 1993 & 181 & [(MgI+CaI+TiI)/3FeI] & thin, thick \\
\\
                       &  &                      & {\it No Kinematics} \\
McWilliam 1995, 1998   & 30 & [(MgI+CaI+TiI)/3FeI] & halo assumed \\
Johnson 2002           & 13 & \nodata & halo assumed \\
Burris et al. 2000     & 15 & \nodata & halo assumed \\
Ivans et al. 2003      &  2 & [(MgI+CaI)/2FeI]      & halo assumed \\
Ryan et al. 1996       & 14 & [(MgI+CaI+TiI)/3FeI] & halo assumed\\
Gratton \& Sneden 1991, 1994 & 10 & [(CaI+(TiI+TiII)/2)/2FeI] & halo assumed \\
Gratton \& Sneden 1988 &  3 & [(MgI+CaI+TiI)/3FeI] & halo assumed \\
\enddata
\tablecomments{References are ordered as included in the dataset for the plots. 
For example stars in Stevens \& Boesgaard were neglected if also in Fulbright.
Four stars in Bensby \etal overlap with Fulbright but they were retained
for the differential abundances discussed in Section~6.3.   Also, four stars
in Burris \etal overlap with McWilliam but are retained because of their
additional LaII abundances.    Stars in Hanson \etal  were neglected
if also in the McWilliam, Burris \etal, or Johnson datasets, even though 
these latter references are without kinematic information.   The data
used in this analysis is available electronically.}
\end{deluxetable}

\clearpage
\begin{deluxetable}{lrrrcccccccccccccccl} 
\tabletypesize{\scriptsize}
\tablecaption{Assembled Literature Data\label{elect}}
\tablewidth{0pt}
\rotate
\tablehead{
\colhead{Name} & \colhead{U} &  \colhead{V} &  \colhead{W}   & 
\colhead{TN} & \colhead{TK} & \colhead{HL} & \colhead{G} & 
\colhead{Fe/H} & \colhead{Mg/Fe} & \colhead{Ca/Fe} & \colhead{Ti/Fe} & 
\colhead{$\alpha$/Fe}  & \colhead{Na/Fe} & \colhead{Ni/Fe} & \colhead{Y/Fe} & 
\colhead{Ba/Fe} & \colhead{La/Fe} & \colhead{Eu/Fe} & \colhead{Source} 
} 
\startdata
 171       &    -1.0 &  161.0 & -23.0  & 0.2 &  0.8  & 0.0  & 0 &  -1.00  &  0.55  &  0.38  &  0.29  &  0.41   &  0.26  &  0.08  & -0.06 &  -0.19 & 200. &   0.34  & Fulbright (2000,2002)         \\
 2413      &  -161.0 & -125.0 & -46.0  & 0.0 &  0.0  & 1.0  & 1 &  -1.96  &  0.25  &  0.26  &  0.20  &  0.24   & -0.33  & -0.08  & -0.51 &  -0.19 & 200. & 100.  & Fulbright (2000,2002)         \\
 3026      &  -144.0 &   -3.0 & -34.0  & 0.0 &  0.0  & 1.0  & 0 &  -1.32  &  0.27  &  0.33  &  0.31  &  0.30   & -0.17  & -0.05  & -0.06 &   0.15 & 200. & 100. & Fulbright (2000,2002)         \\
 19814     &   337.0 &   40.0 &  33.0  & 0.0 &  0.0  & 1.0  & 0 &  -0.71  &  0.07  &  0.01  &  0.08  &  0.05   & -0.48  & -0.13  & -0.12 &   0.17 & 200. & 2000. &  Stephens and Boesgaard (2002) \\
 G082-023  &  -199.0 & -229.0 & -94.0  & 0.0 &  0.0  & 1.0  & 0 &  -3.49  &  0.29  &  0.14  &  0.31  &  0.25  &  100. &  0.45  & 200. &  -0.37 & 200. & 2000. &  Stephens and Boesgaard (2002) \\
 G4-36     &  1000.0 &   0.0  & 1000.0 & 0.0 &  0.0  & 1.0  & 0 &  -1.93  & -0.19  & -0.21  &  0.54  &  0.05   & -0.28  &  0.48  & 200. &  -0.72 & 200. & 2000. &  Ivans et al. (2003)           \\
CS22966012 & 1000.0  &   0.0  & 1000.0 & 0.0 &  0.0  & 1.0  & 0 &  -1.91  & -0.65  & -0.24  &  0.60  & -0.10   & -0.64  &  0.54  & 200. & 333. & 200. & 2000. &  Ivans et al. (2003)           \\
 HR  8885  &   12.3  & 202.1  & -13.5  & 0.8 &  0.2  & 0.0  & 0 &   0.02  &  0.17  &  0.02  &  0.07  &  0.09   &  0.06  &  0.10  & 9998. &  0.04 & 200. &  99.  & Edvardsson et al. (1993)      \\
 HR  8969  &    7.8  & 195.0  & -25.3  & 0.6 &  0.4  & 0.0  & 0 &  -0.17  &  0.17  &  0.06  &  0.06  &  0.10   &  0.10  &  0.05  & -0.02 &  -0.14 & 200. &   0.22  & Edvardsson et al. (1993)      \\
 HD  2615  &   63.0  & 261.0  & -9.1   & 0.8 &  0.2  & 0.0  & 0 &  -0.58  &  0.12  &  0.11  &  0.12  &  0.12   &  0.08  & -0.02  & -0.09 &  -0.19 & 200. & 100.  & Edvardsson et al. (1993)      \\
 HD  6434  &  -74.0  & 158.4  & -6.0   & 0.1 &  0.9  & 0.0  & 0 &  -0.54  &  0.32  &  0.18  &  0.26  &  0.25   &  0.09  &  0.02  &  0.28 &   0.15 & 200. &   0.35  & Edvardsson et al. (1993)      \\
 Scl459    & 1000.0  &   0.0  & 1000.0 & 0.0 &  0.0  & 0.0  & 1 &  -1.66  &  0.36  &  0.15  &  0.10  &  0.20   & -0.33  &  0.11  & -0.05 &   0.33 &  -0.08 &   0.63  & Shetrone et al. (2003)        \\
 Scl479    & 1000.0  &   0.0  & 1000.0 & 0.0 &  0.0  & 0.0  & 1 &  -1.77  &  0.26  &  0.22  &  0.11  &  0.19   & -0.59  & -0.24  & -0.79 &  -0.19 &  -0.35 &   0.25  & Shetrone et al. (2003)        \\
 \nodata  & \nodata & \nodata & \nodata & \nodata & \nodata & \nodata & \nodata & \nodata & \nodata & \nodata  & \nodata & \nodata & \nodata & \nodata & \nodata & \nodata & \nodata & \nodata & \nodata \\
\enddata
\tablecomments{Starname is from the HIP catalogue, unless otherwise noted.  
UVW are Galactocentric velocities; when velocities are not available we
have used U=W=1000.0 and V=0.0, and adopted a Galactic component for that
star.  TN/TK/HL are the membership probabilities in the thin or thick 
disks or halo as calculated here using the velocity ellipsoids in Table~3.   
The G index refers to dwarfs (=0) or giants (=1), when log(gravity)$\le$3.0.
Any abundance ratio over 9.0 is a dummy value.
The full electronic table includes 821 entries. }
\end{deluxetable}

\clearpage
\begin{deluxetable}{lccccccl} 
\footnotesize
\tablecaption{Velocity Ellipsoids\label{vels}}
\tablewidth{0pt}
\tablehead{
\colhead{Gal. component} & \colhead{$\sigma_U$}  & \colhead{$\sigma_V$}  & 
\colhead{$\sigma_W$}  & \colhead{U}  & \colhead{V}  & \colhead{W}  & 
\colhead{Reference} 
} 
\startdata
thin disk  & 44  & 25  & 20  & 0 & 220 & 0 & Dehnen \& Binney 1998 \\
           &     &     &     & 0 & 220 & 0 & Gilmore, Wyse \& Norris 2002 \\
thick disk & 63  & 39  & 39  & 0 & 180 & 0 & Soubiran \etal 2003 \\
halo       & 141 & 106 & 94  & 0 &  0  & 0 & Chiba \& Beers 2000 \\
\enddata
\end{deluxetable}

\end{document}